\documentclass[sn-mathphys]{sn-jnl}

\jyear{2023}%

\theoremstyle{thmstyleone}%
\theoremstyle{thmstyletwo}%

\theoremstyle{thmstylethree}%

\usepackage{tabularx}
\usepackage{caption}
\usepackage{subcaption}
\usepackage{etoolbox}
 
\begin{document}

\title[X-ray polarimetry of the Galactic center]{X-ray polarization evidence for a 200 years-old flare of Sgr~A$^*$}

\author[1]{\fnm{Fr\'ed\'eric} \sur{Marin}}
\author[2,3]{\fnm{Eugene} \sur{Churazov}}
\author[4,2,3]{\fnm{Ildar} \sur{Khabibullin}}
\author[5]{\fnm{Riccardo} \sur{Ferrazzoli}}
\author[6]{\fnm{Laura} \sur{Di~Gesu}}
\author[1]{\fnm{Thibault} \sur{Barnouin}}
\author[5]{\fnm{Alessandro} \sur{Di Marco}}
\author[7,8]{\fnm{Riccardo} \sur{Middei}}
\author[9,3]{\fnm{Alexey} \sur{Vikhlinin}}
\author[5]{\fnm{Enrico} \sur{Costa}}
% ...

\affil[1]{Université de Strasbourg, CNRS, Observatoire Astronomique de Strasbourg, UMR 7550, 67000 Strasbourg, France}
\affil[2]{Max Planck Institute for Astrophysics, Karl-Schwarzschild-Strasse 1, 85741 Garching, Germany}
\affil[3]{Space Research Institute of the Russian Academy of Sciences, Profsoyuznaya Str. 84/32, Moscow 117997, Russia}
\affil[4]{Universit\"ats-Sternwarte, Fakult\"at f\"ur Physik, Ludwig-Maximilians-Universit\"at M\"unchen, Scheinerstr.1, 81679 M\"unchen, Germany}
\affil[5]{INAF Istituto di Astrofisica e Planetologia Spaziali, Via del Fosso del Cavaliere 100, 00133 Roma, Italy}
\affil[6]{ASI - Agenzia Spaziale Italiana, Via del Politecnico snc, 00133 Roma, Italy}
\affil[7]{Space Science Data Center, Agenzia Spaziale Italiana, Via del Politecnico snc, 00133 Roma, Italy}
\affil[8]{INAF Osservatorio Astronomico di Roma, Via Frascati 33, 00078 Monte Porzio Catone (RM), Italy}
\affil[9]{Center for Astrophysics, Harvard \& Smithsonian, 60 Garden St, Cambridge, MA 02138, USA}
% ...

\abstract{The center of the Milky Way Galaxy hosts a $\sim$4 million solar mass black hole (Sgr~A$^*$)  that is currently very quiescent with a luminosity many orders of magnitude below those of active galactic nuclei [1]. Reflection of X-rays from Sgr~A$^*$ by dense gas in the Galactic Center region offers a means to study its past flaring activity on times scales of hundreds and thousands of years [2]. The shape of the X-ray continuum and the strong fluorescent iron line observed from giant molecular clouds in the vicinity of Sgr~A$^*$ are consistent with the reflection scenario [3-5]. If this interpretation is correct, the reflected continuum emission should be polarized [6]. Here we report observations of polarized X-ray emission in the direction of the Galactic center molecular clouds using the Imaging X-ray Polarimetry Explorer (IXPE). We measure a polarization degree of 31\% $\pm$ 11\%, and a polarization angle of $-$48$^\circ$ $\pm$ 11$^\circ$. The polarization angle is consistent with Sgr~A$^*$ being the primary source of the emission, while the polarization degree implies that some 200 years ago the X-ray luminosity of Sgr~A$^*$ was briefly comparable to a Seyfert galaxy.}


\maketitle

\begin{figure}[!h]
\centering
\includegraphics[width=\textwidth]{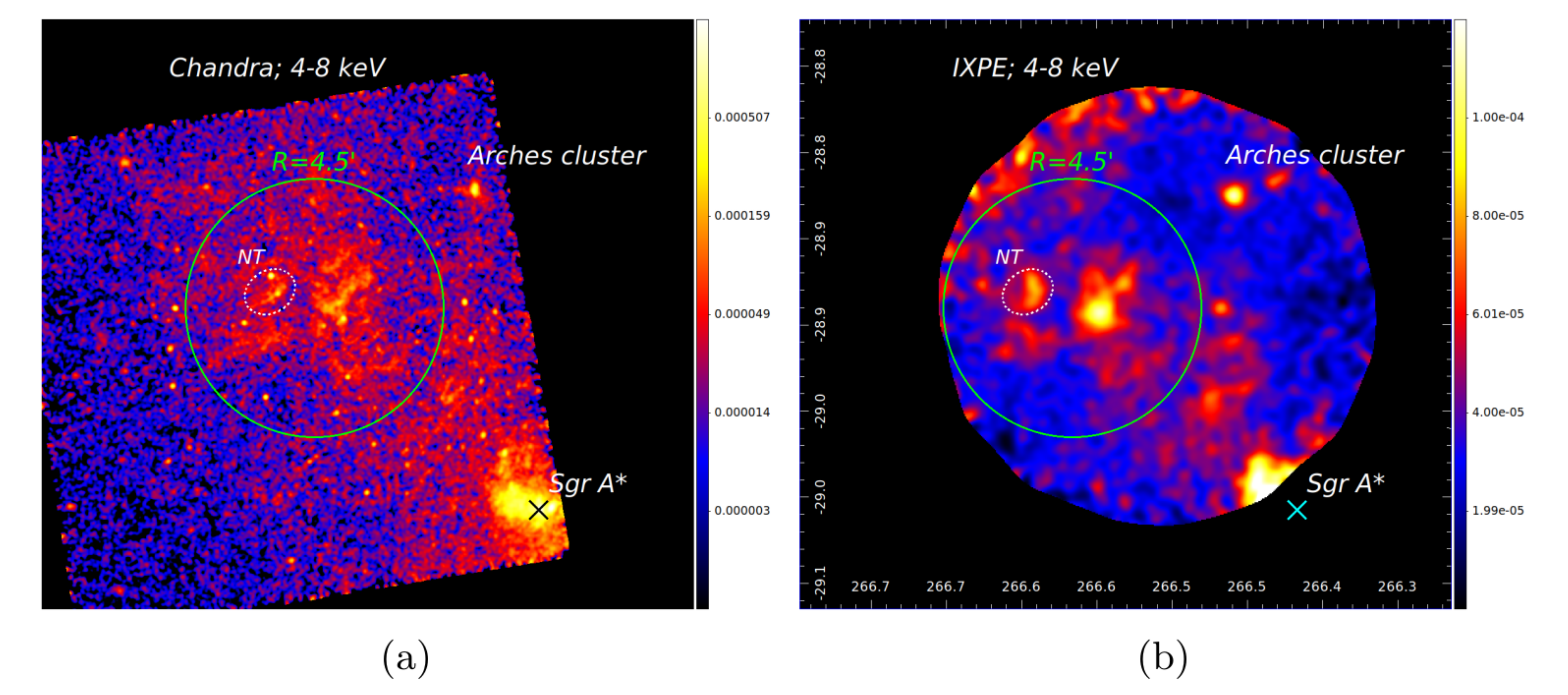}
\caption{\textbf{Quasi-simultaneous Chandra (a) and IXPE (b) 4--8~keV X-ray surface brightness maps of the Galactic center region to the North-East from Sgr~A$^*$.} The maps are in equatorial coordinates, North is up. The IXPE image is in units of ${\rm counts~s^{-1}~arcmin^{-2}~cm^{-2}}$ colour-coded on a linear scale (see the colorbar on the right). For IXPE, the edges of the image, where the exposure (including vignetting effects) drops below 15\% of the maximal value, have been truncated [26]. For Chandra, the image units are the same but a logarithmic scale is used to show both diffuse and compact sources.   The positions of the supermassive black hole Sgr~A$^*$ and the Arches star cluster are labeled. The area where reflection emission is strong in the Chandra data is shown using a green $4.5'$ radius circle. This circle was used to extract I, Q, and U spectra from the IXPE data. The same region was used for {Chandra} and {XMM-Newton} spectral analysis. The white dashed ellipse shows the location of a bright non-thermal (NT) source G0.13-0.11 [11] that was excluded for the spectra extraction.}
\label{Fig:IXPE_Chandra_map}
\end{figure}

\begin{figure}[!h]
\centering
\includegraphics[angle=0,trim= 0mm 4cm 0mm 1cm,width=1\columnwidth]{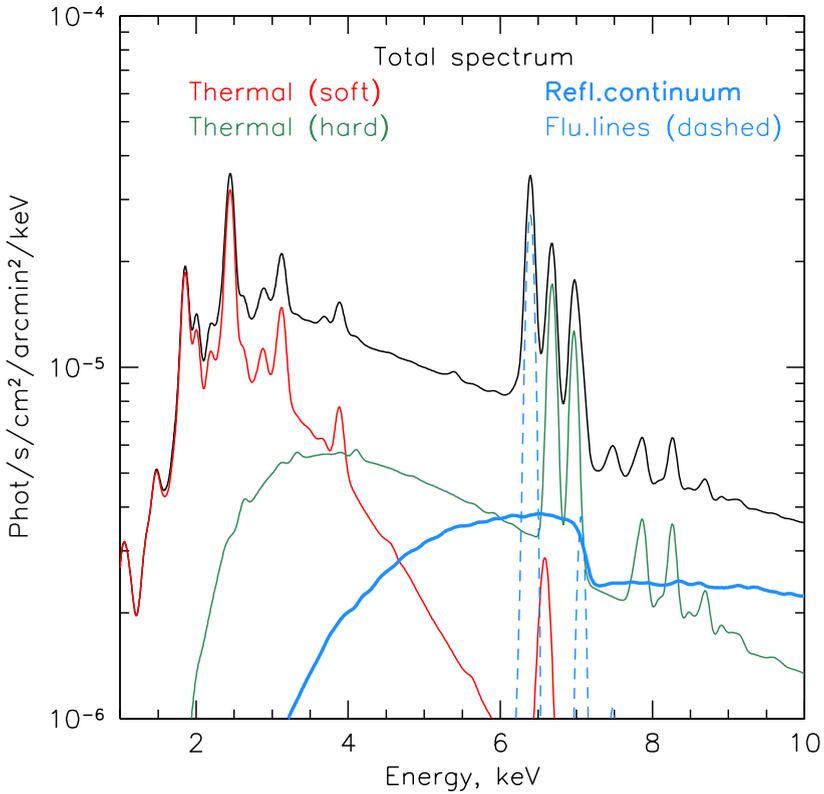}
\caption{\textbf{An illustration of the spectral model used to approximate X-ray spectra extracted from the reference region.} For clarity, only two ``thermal" components are shown in the plot. The red and the green curves are those thermal components, detailed in [45,46]. The hotter (green) of the models, having prominent lines at 6.7 and 6.97 keV, contributes significantly to the 4--8 keV band. The blue curves show the two components of the reflected emission. Namely, the dashed blue curves show the fluorescent lines of iron (K$\alpha$ line at 6.4 keV and K$\beta$ line at 7.06 keV), while the thick blue line shows the scattered continuum. Only the latter component is polarized and used to fit the Q and U spectra measured by IXPE. The black line shows the sum of all components. In order to show more clearly the components of the physical model in units of ${\rm photons\,s^{-1}\,cm^{-2}\,arcmin^{-2}\,keV^{-1}}$ the spectra were convolved with a Gaussian, which is narrower than the energy resolution of the IXPE, Chandra, and XMM-Newton detectors.}
\label{fig:spectral_model_demo}
\end{figure}

IXPE observed the molecular complex Sgr~A, the X-ray brightest group of reflection clouds near Sgr~A$^*$ [7], for an exposure of 0.93 megaseconds after filtering in February and March 2022. Contemporaneous Chandra X-ray Observatory observations with high angular resolution were acquired to identify regions where the reflected emission is strong and to extract spectra from these regions. Fig.~\ref{Fig:IXPE_Chandra_map} shows the X-ray surface brightness maps in the 4--8 keV band obtained. We excluded X-rays below 4 keV since they arise from unpolarized plasma emission [8-10], see Fig.~\ref{fig:spectral_model_demo}. See the Methods section for details on the observations and data reduction.

Based on the Chandra image only, we selected a circular extraction region with a radius $4.5'$ centered on the Sgr A complex that contains numerous hot spots of reflected emission (i.e. scattering clouds). This region was selected to collect as many `reflected' photons as possible while avoiding the edges of the IXPE detectors. The region includes most areas of elevated reflected emission. The region excludes areas affected by very bright sources that enter the IXPE field-of-view as a result of the pointing dithering pattern and a bright $1'$ region associated with the non-thermal object G0.13-0.11 [11] to avoid possible contamination of the polarised signal.

\begin{figure}[!h]
\centering
\includegraphics[trim=1cm 5cm 0cm 2cm,width=\textwidth]{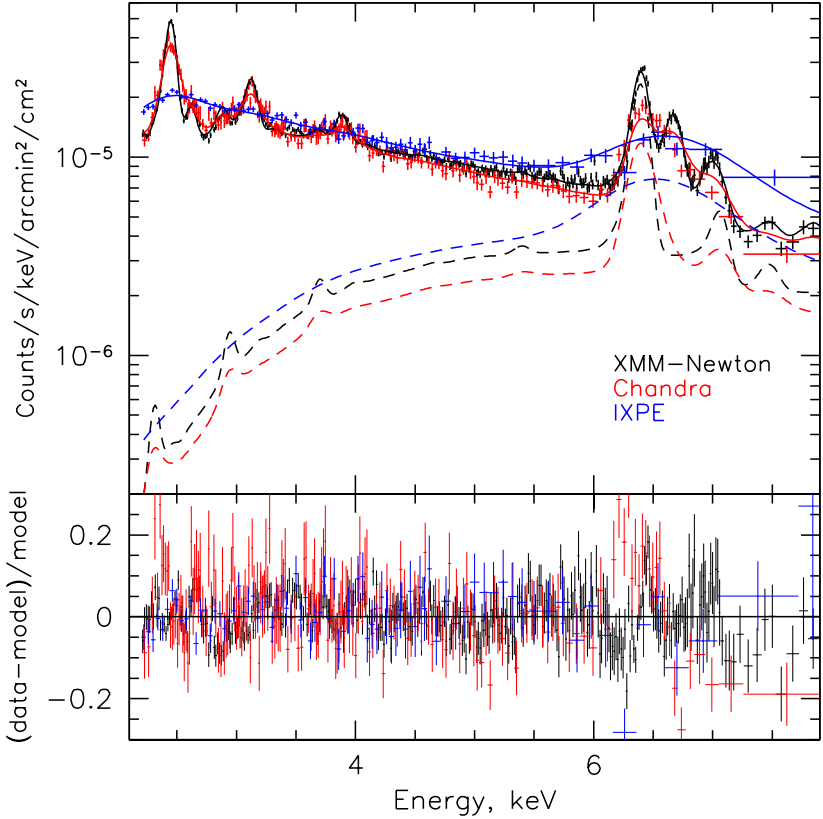}
\caption{\textbf{Spectra of the X-ray emission extracted from the circular region shown in Fig.~\ref{Fig:IXPE_Chandra_map} obtained in Chandra (red), IXPE (blue) and archival XMM-Newton (black) observations, after division by energy-dependent effective area of each telescope.} Errorbars correspond to 1$\sigma$ (68\% confidence intervals). The solid lines correspond to the best-fitting models (three thermal components and the reflection emission, see Methods) convolved with the telescopes' spectral responses, resulting in strong smearing of the emission lines in the IXPE data. Parameters of the thermal components were fixed among all three datasets, and only the normalization of the reflection component was allowed to vary between them. The dashed lines show 
the best-fitting contribution of the reflected emission to the total spectrum for each of the three instruments. The difference in the normalization of the reflected emission ($\sim 30$\%) could be caused by the time variability and the different energy-dependent efficiencies of the three telescopes.
As demonstrated by the residuals shown in the second panel, the model provides 
a reasonable approximation of the data (at the level of $\sim$20\%) even although with high statistics of the XMM-Newton and Chandra data a few wiggles near the bright emission lines are visible, which are plausibly caused by slight gain variations among the data sets.}
\label{Fig:IXPE_Chandra_spectra}
\end{figure}

Fig.~\ref{Fig:IXPE_Chandra_spectra} shows good agreement between the spectra extracted from IXPE, Chandra, and archival XMM-Newton data for the selected region.  A bump, associated with iron lines at 6.4 and 6.7~keV, is clearly seen, even with IXPE's lower spectral resolution. The model based on XMM-Newton and Chandra data well describes the IXPE spectrum, given the current level of cross-calibration uncertainties between the instruments.

The spectral model is built from several components: thermal and reflected emission, fluorescent lines, and scattered continuum (see Methods). All of them, except for the component representing the reflected emission from the molecular cloud, are assumed to be constant in time, i.e. the same for all three datasets. The reflected emission is a combination of Compton scattered and photo-absorbed continuum with fluorescent emission lines calculated self-consistently from the reflection spectrum [12]. Only the continuum part of the reflected spectrum is expected to be polarized [9,10,13]. Hence, the X-ray polarization (described by the Stokes parameters Q and U measured by IXPE) is related to the reflection continuum I$_{\rm refl,c}$, and not to the total observed intensity I [14-15]. Polarization degree is independent of photon energy in the single-scattering scenario, so the spectral shape of Q and U follow that of I$_{\rm refl,c}$. 

\begin{figure}[!h]
\centering
\includegraphics[angle=0,trim= 0mm 4cm 0mm 1cm,width=0.65\columnwidth]{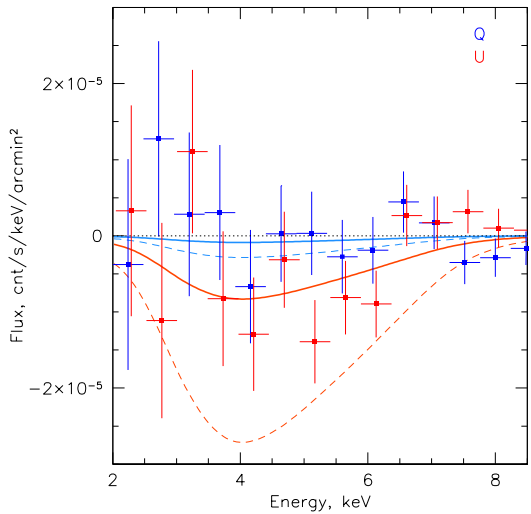}
\caption{\textbf{Stokes parameters Q (blue) and U (red) spectra extracted from the circular region shown in Fig.~\ref{Fig:IXPE_Chandra_spectra}.} Errorbars correspond to 1$\sigma$ (68\% confidence intervals). For clarity, both the Q and U data points are slightly shifted in energy. The solid blue and red lines represent the best fitting reflection model (the Compton-scattered continuum component only) to the Q and U spectra with a degree of polarization 31\% and the polarization angle $-48^\circ$. The model has been convolved with the IXPE spectral response. The thinner dashed lines are representative of the maximum Q and U that one could expect for 90$^\circ$ scattering that would lead to a 100\% polarized reflected continuum.}
\label{Fig:IXPE_Q_U}
\end{figure}

\begin{figure}[!h]
\centering
\includegraphics[trim= 0mm 0cm 0mm 1cm,width=1\columnwidth]{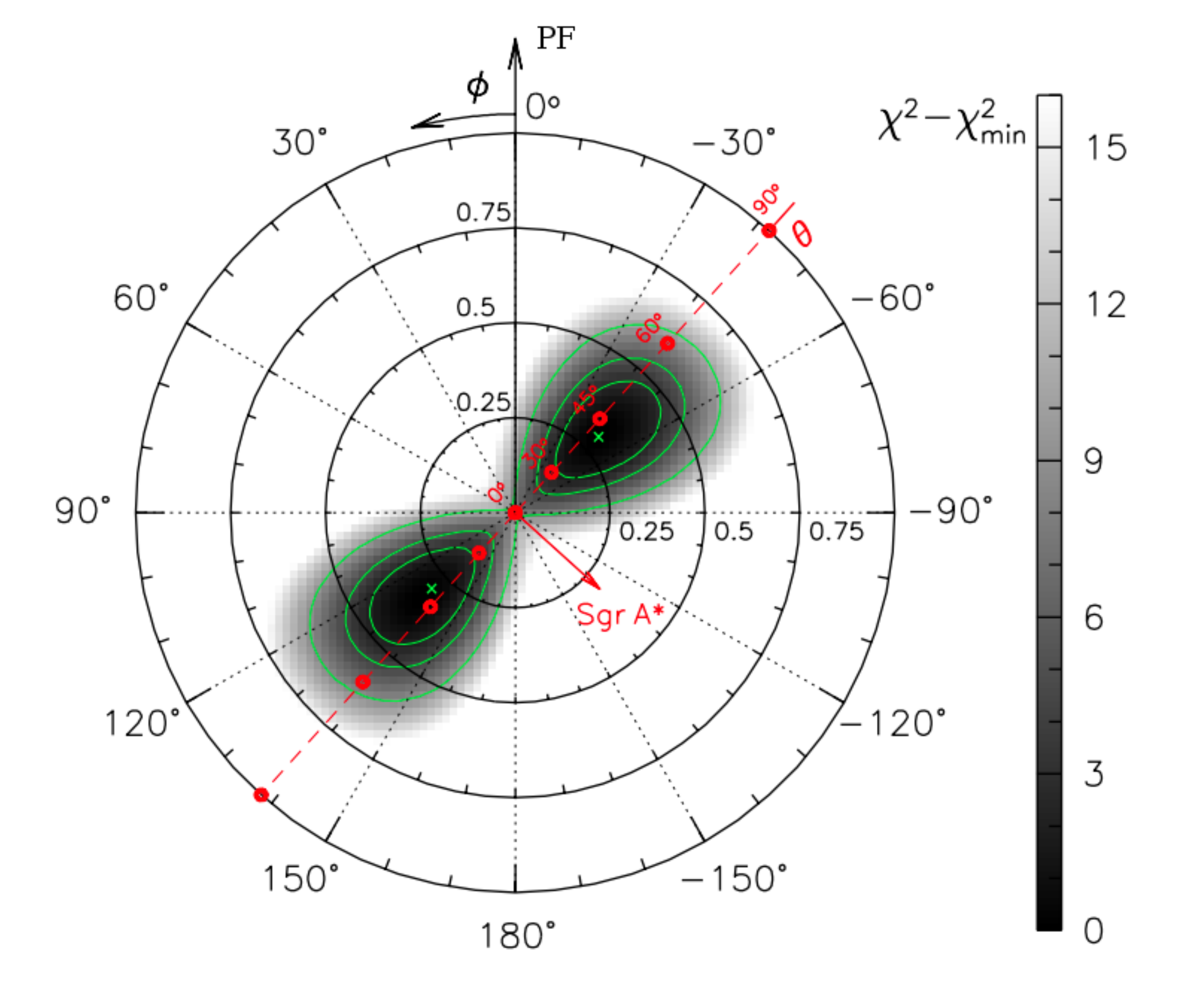}
\caption{\textbf{Map of the $\chi^2-\chi^2_{\min}$ statistic for the fit of the Stokes parameters Q and U spectra}.,We used the reflection continuum model in polar coordinates with radius equal to the polarization fraction (PF) and the azimuthal angle showing the position angle of the electric field vector ($\phi$). The minimal value of the statistic $\chi^2_{\min}$ corresponds to the best fitting values $\phi=-48^\circ$ and $P=31$\% (marked as green crosses), and the contours show 68, 90 and 99\% confidence levels. The hypothesis that Sgr~A$^*$ is the primary source of illuminating X-ray flux, implies the polarization angle $\phi=-42^\circ$ for the center of the region used for extraction of the Q and U spectra, as marked with the red dashed line. The circles on this line depict expected polarization degrees for the scattering angle changing from 0(180) to 30(150), 45(135), 60(120), and 90 degrees.}
\label{Fig:IXPE_Flare}
\end{figure}

Fig.~\ref{Fig:IXPE_Q_U} shows the observed Q and U spectra. The Q values are close to zero, while U is mostly negative, suggesting a polarization angle close to $-45^\circ$. The best-fitting reflection model is shown with solid lines. Here, we have explicitly assumed that all emission from the selected region has the same polarisation degree and angle, which is valid if the illuminated group of clouds is compact in 3D. With these simplifying assumptions, our model has only two free parameters derived from the IXPE data: degree of polarization $P=31\pm 11$\% and polarisation angle $\phi=-48^\circ\pm 11^\circ$ (uncertainties are 68\% confidence level, see Fig.~\ref{Fig:IXPE_Flare}; $\phi$ is anti-clockwise from North in the equatorial coordinate system per IAU convention). As explained in Methods, the absolute normalization of the reflected component has $\sim$30\% uncertainty. Therefore, the constraints on the polarization degree are a multiplicative combination of statistical and systematic uncertainties: $P=(0.31\pm 0.11)_{\rm stat} \times (1.0 \pm 0.3 )_{\rm sys}$.  Even with this extra uncertainty, it is clear that the observed polarization degree is significantly smaller than 100\%. The systematic uncertainty does not affect the significance of the polarized signal detection, which corresponds to $\sim$ 2.8 standard deviations.

The measured polarization angle $\phi$ is consistent with the hypothesis that Sgr~A$^*$ is the primary source (see Fig.~\ref{Fig:IXPE_Flare}). For a cloud located at the center of our extraction region, the line orthogonal to the direction towards Sgr~A$^*$ corresponds to $\phi = -42^\circ$. This is within the uncertainties of our measurement ($\phi=-48^\circ\pm 11^\circ$), so fixing the measured polarization angle to $-42^\circ$ does not affect the errors on the degree of polarisation.

The degree of polarisation $P$ of the reflected continuum is directly related to the scattering angle $\theta$ in the single scattering approximation, $P=(1-\mu^2)/(1+\mu^2)$, where $\mu=\cos\theta$. Any value of $P$ produces two solutions, $\theta$ and $\pi-\theta$. For $P=31\pm 11$\%, the solutions are $ 43 \substack{+7 \\ -8} $ and  $ 137 \substack{-7 \\ +8} $ degrees. A similar level of uncertainty is caused by the $\sim 30\%$ systematic error in the value of $P$. These angles fix the scattering geometry, the smaller value corresponds to clouds between us and the primary source and the larger to clouds beyond the source (see Methods).

For a given scattering angle $\theta$, one can also calculate the age of the flare $t_{\rm flare}$, i.e. the time delay associated with the propagation of X-rays from the primary source to the cloud and then to the observer. Adopting a projected distance between Sgr~A$^*$ and the scattering clouds of 25~pc, i.e. the distance between the SMBH and the centre of our extraction zone in the sky plane, the two solutions for $\theta$ translate into two $t_{\rm flare}$ values: 
$ 33 \substack{+6 \\ -7} $ and  $ 205 \substack{-30 \\ +50}$ years, respectively. Again, the systematic uncertainties are comparable to the quoted statistical errors (see Methods). From the point of view of polarization properties, both solutions are equivalent. However, the `older' flare is far more plausible. First of all, for the cloud located well behind the primary source, the propagation speed of the light front of the flare approaches $c/2$ [16], where $c$ is the speed of light, while for the cloud in front of the primary source, this speed is always greater than $c$. Thus, for a short flare, a cloud located behind Sgr~A$^*$ will remain bright longer than the same cloud in front of it, and, therefore, on average, the chances to spot it in a bright phase are higher. Secondly, the `younger' flare from Sgr~A$^*$ would have been observed directly. Indeed, the Advanced Satellite for Cosmology and Astrophysics (ASCA) was active 30 years ago; the GMCs were bright at the time [5,17], while Sgr~A$^*$ was not. The Sgr~B2 cloud, which in the sky plane is much further away from Sgr~A$^*$ than the area covered by IXPE, was bright too. This fact strongly suggests that the flare is significantly older than 30 years, or that there were multiple flares. Here, the former scenario is adopted and is consistent with the range of values found in the literature (100 -- 500 years, depending on the methods and Galactic center regions probed [18-21]).

Our work presents the missing piece of evidence that X-rays from the giant molecular clouds are due to reflection of an intense, yet short-lived flare produced at or nearby Sgr~A$^*$. These results can further constrain the past activity of the Galactic center. The observational data provide a flare fluence (product of the luminosity $L_X$ and duration $\Delta t$ of the flare) of a few $10^{47}$~erg [12,22]. Based on previous studies of the X-ray variability of small molecular clouds in reflected emission, the duration of the flare is $\Delta t < 1.6\,{\rm yr}$ [22]. Existing observational data [12,22], including those of IXPE (see Methods), suggest the broad-band (1--100~keV) flare luminosity was in the range from a few $10^{39}$ to $\sim 10^{44}$~erg\,s$^{-1}$, i.e. comparable to the X-ray luminosity of Seyfert galaxies [23,24]. The lower limit comes from the brightest levels of the surface brightness observed over years (see Methods), while the upper limit corresponds to the Eddington luminosity of Sgr~A*, which could be exceeded in some scenarios. Given the fluence constraint, the lowest luminosity requires a 1--2~year-long flare, while the highest luminosity needs only an hour-long outburst. 

Potential origins of flares with the needed fluence [25] include accretion induced collapse of white dwarfs, sub-luminous gamma-ray bursts, tidal disruption events, or transient accretion onto SMBHs not related to tidal disruption. Classes of transients that can be excluded because their luminosities fall short by a factor of 10--100 include giant flares from soft gamma-ray repeaters (magnetars), type II supernova breakouts, and major outbursts of ultra-luminous X-ray sources. The second line of reasoning could come from the comparison of the expected event rates, bearing in mind that the fluence was calculated for the adopted distance between the clouds and Sgr~A*, and, therefore, the assumption that the source is further away would boost the required fluence. These considerations favor Sgr~A* as the prime candidate, especially considering the consistency of its position with the measured polarization degree, although other scenarios are not completely eliminated. 

Detection of the faint and diffuse X-ray emission due to reflection from molecular clouds in the Galactic Center region remains a challenging exercise. Current IXPE data demonstrate this goal is within the reach of IXPE, paving a way for future, even longer observations that will enable more detailed analysis, including spatially resolved polarisation, to verify the single flare scenario, and the determination of the intrinsic polarization of the flares to probe the origin of the radiation and constraints on the multiple-flare scenarios.

\backmatter

\bmhead{Acknowledgments}
The Imaging X ray Polarimetry Explorer (IXPE) is a joint US and Italian mission.  The US contribution is supported by the National Aeronautics and Space Administration (NASA) and led and managed by its Marshall Space Flight Center (MSFC), with industry partner Ball Aerospace (contract NNM15AA18C).  The Italian contribution is supported by the Italian Space Agency (Agenzia Spaziale Italiana, ASI) through contract ASI-OHBI-2017-12-I.0, agreements ASI-INAF-2017-12-H0 and ASI-INFN-2017.13-H0, and its Space Science Data Center (SSDC) with agreements ASI-INAF-2022-14-HH.0 and ASI-INFN 2021-43-HH.0, and by the Istituto Nazionale di Astrofisica (INAF) and the Istituto Nazionale di Fisica Nucleare (INFN) in Italy.  This research used data products provided by the IXPE Team (MSFC, SSDC, INAF, and INFN) and distributed with additional software tools by the High-Energy Astrophysics Science Archive Research Center (HEASARC), at NASA Goddard Space Flight Center (GSFC).
F.M. is grateful to the Astronomical observatory of Strasbourg, the CNRS and the University of Strasbourg under whose benevolence this paper was written. I.K. acknowledges support by the COMPLEX project from the European Research Council (ERC) under the European Union’s Horizon 2020 research and innovation program grant agreement ERC-2019-AdG 882679. P.O.P. acknowledges financial support from the French National Program of High Energy (PNHE/CNRS) and from the French national space agency (Centre National d’Etudes Spatiales -- CNES). I.A. acknowledges financial support from the Spanish "Ministerio de Ciencia e Innovaci\'on" (MCIN/AEI/ 10.13039/501100011033) through the Center of Excellence Severo Ochoa award for the Instituto de Astrof\'isica de Andaluc\'ia-CSIC (CEX2021-001131-S), and through grants PID2019-107847RB-C44 and PID2022-139117NB-C44. A.I. acknowledges support from the Royal Society. A.V., W.F., and R.K. acknowledge support from NASA Grant GO1-22136X, the Smithsonian Institution, and the Chandra High Resolution Camera Project through NASA contract NAS8-03060. C.-Y. Ng is supported by a GRF grant of the Hong Kong Government under HKU 17305419.
Finally, the authors are grateful to the anonymous referees for the evaluation of our paper and for the constructive critics.

\bmhead{Author contributions}
F. Marin led the IXPE observation, contributed to the analysis and led the writing of the paper. E. Churazov, I. Khabibullin, R. Ferrazzoli, L. Di Gesu, T. Barnouin, A. Di Marco, R. Middei, E. Costa, P. Soffitta, F. Muleri, R. Sunyaev and P. Kaaret contributed to the IXPE analysis, discussion and writing of the paper. A. Vikhlinin, W. Forman and R. Kraft provided and reduced the Chandra data used in this paper. S. Bianchi, D. Immacolata, P.-O. Petrucci and T. Enoto contributed with discussion and parts of the paper. The remaining authors are part of the IXPE team whose significant contribution made the satellite and the Galactic center observation possible. 

\bmhead{Author information}
The main author can be contacted using his official e-mail address : frederic.marin@astro.unistra.fr

\bmhead{Competing interest}
Authors declare that they have no competing interests.

\bmhead{Data availability}
The IXPE data that support the findings of this study are freely available in the HEASARC IXPE Data Archive (\url{https://heasarc.gsfc.nasa.gov/docs/ixpe/archive/}). The XMM-Newton data can be found on the same website, while the Chandra data will be public after one year from the observation date.

\bmhead{Code availability}
The analysis and simulation software \textit{ixpeobssim} developed by IXPE collaboration and its documentation is available publicly through the web-page \url{https://ixpeobssim.readthedocs.io/en/latest/?badge=latest.494}.
XSPEC is distributed and maintained under the aegis of the HEASARC and can be downloaded as part of HEAsoft from
\url{http://heasarc.gsfc.nasa.gov/docs/software/lheasoft/download.html}.

\begin{figure}[!h]
\centering
\includegraphics[width=\textwidth]{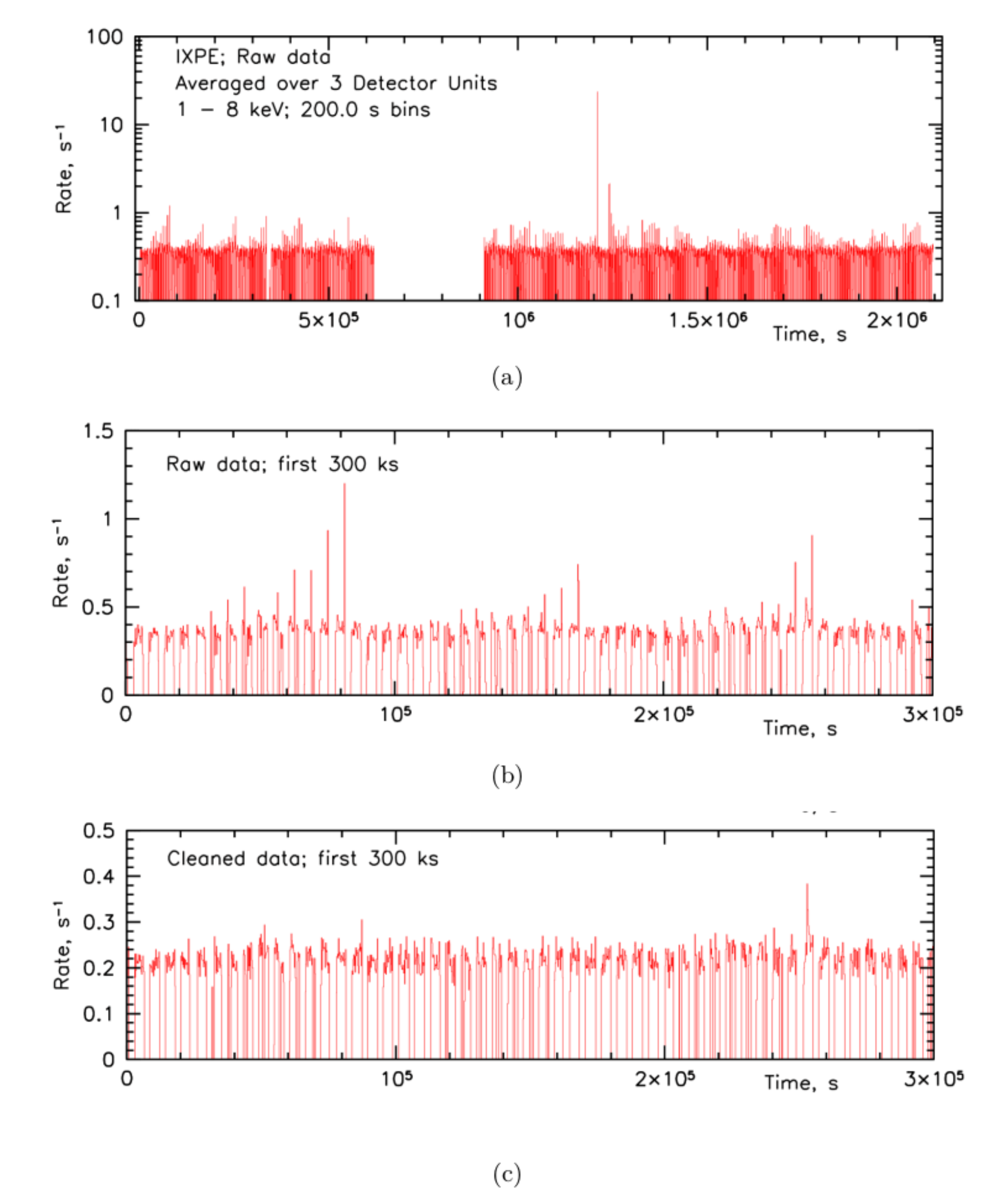}
\caption{\textbf{Illustration of the data cleaning prior to the imaging and spectral analysis.} Panel (a) shows the count rate in the 2--8 keV band (sum of three DUs, 200 seconds time bins) in the original data set that spans 2 million seconds. The two most intense spikes are associated with a geomagnetic storm. To illustrate less prominent variations, panel (b) shows a 300~kiloseconds-long portion of the same light curve that feature a number of smaller amplitude quasi-regular spikes (notice that the vertical scale has changed) that are mostly due to the South Atlantic Anomaly. The gaps in the light curves correspond to moments when the Galactic center was obscured by the Earth. Finally, panel (c) shows the count rate for the data cleaned from spikes and individual events that most plausibly are due to detector background rather than X-ray photons. The overall count rate in the cleaned data is almost a factor of two lower than in the original data.}
\label{Fig:lc}
\end{figure}

\begin{table}[htbp]
\begin{center}
\caption{Observations used in this work for background extraction.
\label{tab:bkgobservations}}
\begin{tabular}{cccccc}
\hline\hline
\textbf{Mission}                   & \textbf{Obs.ID}                 & \textbf{Start date}                  & \textbf{Effective exposure}     & \textbf{Notes} 	 \\
			                 & 	                      & (\textbf{YYYY-MM-DD)}                & \textbf{(kiloseconds)}  	             &                  \\
\hline
IXPE 	                  & 1003701                & 2022-05-04                  & 96.7	                  & Mrk421          \\
IXPE 	                  & 1003801                & 2022-06-04                  & 96.0	                  & Mrk421          \\
IXPE 	                  & 1004501                & 2022-03-08                  & 100.4	              & Mrk501          \\
IXPE 	                  & 1004701                & 2022-07-09                  & 97.8	                  & Mrk501          \\
IXPE 	                  & 1006301                & 2022-05-06                  & 390.9	              & BL Lac          \\
IXPE 	                  & 1006301                & 2022-05-06                  & 116.9	              & BL Lac          \\
IXPE 	                  & 1003501                & 2022-07-12                  & 771.8	              & Circinus          \\
IXPE 	                  & 1005701                & 2022-06-12                  & 264.2	              & 3C279          \\
\hline
\end{tabular}
\end{center}
\end{table}

\begin{table}
\centering
\caption{The three more intense sources and their re-scaled flux in the extraction region we used in our paper (see Fig.~\ref{Fig:IXPE_Chandra_map}). Fluxes are in erg~s$^{-1}$~cm$^{-2}$. \label{Tab:Sources}}
\begin{tabularx}{\textwidth}{l X X X X} 
\hline\hline
\textbf{Source} & \textbf{Flux} & \textbf{Angular sep. ($^\circ$)} & \textbf{Suppression factor} & \textbf{Re-scaled flux}  \\
\hline
2CXO J174621.1-284343 & 5.033E-11 & 0.22 & 245 & 1.4E-13 \\
2CXO J174451.6-292042 & 3.949E-11 & 0.52 & 752 & 3.4E-14 \\
2CXO J174445.4-295045 & 3.387E-11 & 0.964 & 1120 & 2.0E-14 \\
\hline
\end{tabularx}
\end{table}

\clearpage

\subsection*{References}

[1] Reinhard Genzel, Frank Eisenhauer, and Stefan Gillessen. The Galactic Center massive black hole and nuclear star cluster. Reviews of Modern Physics, 82(4):3121–3195, October 2010. \newline \newline
[2] L. A. Vainshtein and R. A. Syunyaev. The K-Alpha Lines in the Background X-Ray Spectrum and the Interstellar Gas in Galaxies. Soviet Astronomy Letters, 6:353–356, June 1980. \newline \newline
[3] R. A. Sunyaev, M. Markevitch, and M. Pavlinsky. The Center of the Galaxy in the Recent Past: A View from GRANAT. Astrophysical Journal, 407:606, April 1993. \newline \newline
[4] M. Markevitch, R. A. Sunyaev, and M. Pavlinsky. Two sources of diffuse X-ray emission from the Galactic Centre. Nature, 364(6432):40–42, July 1993. \newline \newline
[5] Katsuji Koyama and et al. ASCA View of Our Galactic Center: Remains of Past Activities in X-Rays? Publications of the Astronomical Society of Japan, 48:249–255, April 1996. \newline \newline
[6] E. Churazov, R. Sunyaev, and S. Sazonov. Polarization of X-ray emission from the Sgr B2 cloud. Monthly Notices of the Royal Astronomical Society, 330(4):817–820, March 2002. \newline \newline
[7] Ildar Khabibullin, Eugene Churazov, and Rashid Sunyaev. SRG/eROSITA view of X-ray reflection in the Central Molecular Zone: a snapshot in September-October 2019. Monthly Notices of the Royal Astronomical Society, 509(4):6068–6076, February 2022 \newline \newline
[8] M. Revnivtsev and et al. Discrete sources as the origin of the Galactic X-ray ridge emission. Nature, 458(7242):1142–1144, April 2009. \newline \newline
[9] L. Di Gesu and et al. Prospects for IXPE and eXTP polarimetric archaeology of the reflection nebulae in the Galactic center. Astronomy and Astrophysics, 643:A52, November 2020. \newline \newline
[10] R. Ferrazzoli and et al. Prospects for a polarimetric mapping of the Sgr A molecular cloud complex with IXPE. Astronomy and Astrophysics, 655:A108, November 2021. \newline \newline
[11] Q. Daniel Wang, Fangjun Lu, and Cornelia C. Lang. X-Ray Thread G0.13-0.11: A Pulsar Wind Nebula? Astrophysical Journal, 581(2):1148–1153, December 2002. \newline \newline
[12] E. Churazov, I. Khabibullin, G. Ponti, and R. Sunyaev. Polarization and long-term variability of Sgr~A$^*$ X-ray echo. Monthly Notices of the Royal Astronomical Society, 468(1):165–179, June 2017. \newline \newline
[13] F. Marin, F. Muleri, P. Soffitta, V. Karas, and D. Kunneriath. Reflection nebulae in the Galactic center: soft X-ray imaging polarimetry. Astronomy and Astrophysics, 576:A19, April 2015. \newline \newline
[14] F. Kislat, B. Clark, M. Beilicke, and H. Krawczynski. Analyzing the data from X-ray polarimeters with Stokes parameters. Astroparticle Physics, 68:45–51, August 2015. \newline \newline
[15] Ildar Khabibullin, Eugene Churazov, and Rashid Sunyaev. Impact of intrinsic polarization of Sgr~A$^*$ historical flares on (polarization) properties of their X-ray echoes. Monthly Notices of the Royal Astronomical Society, 498(3):4379–4385, September 2020. \newline \newline
[16] R. Sunyaev and E. Churazov. Equivalent width, shape and proper motion of the iron fluorescent line emission from molecular clouds as an indicator of the illuminating source X-ray flux history. Monthly Notices of the Royal Astronomical Society, 297(4):1279–1291, July 1998. \newline \newline
[17] Shigeo Yamauchi, Yoshitomo Maeda, and Katsuji Koyama. ASCA observation of the Galactic Center. Advances in Space Research, 19(1):63–70, May 1997. \newline \newline
[18] Syukyo G. Ryu, Katsuji Koyama, Masayoshi Nobukawa, Ryosuke Fukuoka, and Takeshi Go Tsuru. An X-Ray Face-On View of the Sagittarius B Molecular Clouds Observed with Suzaku. Publications of the Astronomical Society of Japan, 61:751, August 2009. \newline \newline
[19] Syukyo Gando Ryu and et al. X-Ray Echo from the Sagittarius C Complex and 500-year Activity History of Sagittarius A*. Publications of the Astronomical Society of Japan, 65:33, April 2013. \newline \newline
[20] R. Terrier and et al. Fading Hard X-ray Emission from the Galactic Center Molecular Cloud Sgr B2. Astrophysical Journal, 719(1):143–150, August 2010. \newline \newline
[21] D. Chuard and et al. Glimpses of the past activity of Sgr~A$^*$ inferred from X-ray echoes in Sgr C. Astronomy and Astrophysics, 610:A34, February 2018. \newline \newline
[22] E. Churazov, I. Khabibullin, R. Sunyaev, and G. Ponti. Can Sgr~A$^*$ flares reveal the molecular gas density PDF? Monthly Notices of the Royal Astronomical Society, 471(3):3293–3304, November 2017. \newline \newline
[23] A. Ptak. Low-luminosity AGN and normal galaxies. In Nicholas E. White, Giuseppe Malaguti, and Giorgio G. C. Palumbo, editors, X-ray Astronomy: Stellar Endpoints, AGN, and the Diffuse X-ray Background, volume 599 of American Institute of Physics Conference Series, pages 326–335, December 2001. \newline \newline
[24] Eve L. Halderson, Edward C. Moran, Alexei V. Filippenko, and Luis C. Ho. The Soft X-Ray Properties of Nearby Low-Luminosity Active Galactic Nuclei and their Contribution to the Cosmic X-Ray Background. Astronomical Journal, 122(2):637–652, August 2001. \newline \newline
[25] Alicia Soderberg and et al. The Dynamic X-ray Sky of the Local Universe. In astro2010: The Astronomy and Astrophysics Decadal Survey, volume 2010, page 278, January 2009. \newline \newline
[26] Martin C. Weisskopf and et al. The Imaging X-Ray Polarimetry Explorer (IXPE): Pre-Launch. Journal of Astronomical Telescopes, Instruments, and Systems, 8(2):026002, April 2022.

\clearpage

\section*{Methods}

\subsection*{IXPE and Chandra observation}

The molecular complex Sgr~A is currently the brightest collection of reflection clouds in X-ray emission around Sgr~A$^*$ [27]. This region, centered on Right Ascension 266.51$^\circ$ and Declination $-28.89^\circ$ in equatorial coordinates, was pointed by IXPE [28] with a total integration time of 0.93 megaseconds (324 kiloseconds from February 27 to March 6, 2022, and 639 kiloseconds from March 10 to March 24, 2022), after filtering for episodes of enhanced instrumental background.

In addition to the X-ray polarimetric data, a few quasi-simultaneous observations with the Chandra X-ray observatory were acquired from February 26 to March 11, 2022 using the ACIS-I detector (OBSIDs 24373, 24820, and 26353), which cover the 0.3--10.0 keV band with a spectral resolution of $\Delta E \sim$ 0.3~keV near $E=6$~keV and arcsecond angular resolution. The Chandra high angular resolution data were essential to identify regions where the reflected emission is strong and to extract spectra from these regions.

\subsection*{IXPE data reduction}

The IXPE observatory, as described in detail in [28], includes three identical X-ray telescopes, each comprising an X-ray mirror module assembly (provided by NASA) and a polarization-sensitive Gas Pixel Detector (GPD, provided by the Italian side of the collaboration), to offer spatially resolved X-ray polarimetry in the 2--8 keV energy band. At the Science Operations Center (SOC, at the NASA Marshall Space Flight Center), a software pipeline developed jointly by ASI and NASA processes the relevant science, engineering and ancillary data, estimates the photoelectron emission direction (and hence the polarization), position, and energy of each event after applying corrections for charging effects of the Gas Electron Multiplier (GEM), detector temperature, and gain non-uniformity. The use of the IXPE on-board calibration sources [29] allows correction for time-and-spatially-dependant gain variations of the detectors. The in-flight calibration measurements provide the best knowledge of the gain of the detectors at the time of the observation, and hence the correct energy of each photon, needed to correct the Stokes parameters for the presence of spurious polarization, and to use the correct value of the modulation factor (the modulation amplitude for a 100\% linearly polarized radiation [28]). The removal of spurious polarization is achieved using the algorithm of [30]. \\

The output of this pipeline processing is an event file in FITS format for each of the three IXPE detector units that, in addition to the typical information related to spatially resolved X-ray astronomy, contains the event-by-event Stokes parameters [see 31] from which the polarization of the radiation can be derived. The data products are archived at the High-Energy Astrophysics Science Archive Research Center (HEASARC, at the NASA Goddard Space Flight Center), for use by the international astrophysics community. These level-2 event files were cleaned from additional environmental and background contamination [32].\\

In the case of the observation of the Galactic center, the observation window corresponded to a period of increased solar activity which led to an increase of atmospheric noise (particle background). In addition, such activity tends to extend the South Atlantic Anomaly (SAA), above which the detectors are usually turned off to prevent observation in a highly particle polluted environment. However, as can be seen on Fig.~\ref{Fig:lc} (panel a), several spikes were recorded in the light-curve right before the spacecraft entry in the SAA. Using the v26.0.0 ixpeobssim tool (available here :  \url{https://ixpeobssim.readthedocs.io/en/latest/}) [33], we removed the count spikes associated with Earth occultation periods and above the extended SAA. Some spurious events attributed to geomagnetic storm events were also flagged and excised. New good time intervales (GTIs) were then computed using these prolonged epochs and allowed event filtering based on the presence of unaccounted atmospheric noise. After GTI correction (Fig.~\ref{Fig:lc}, panel c), the livetime (or Total Good Time) is 0.93~Ms, which represent 96.7\% of the livetime before filtering.

\subsection*{IXPE systematic effects}
All systematic errors are well below the polarization degree and angle observed from the $4.5'$ extraction circle we used in this paper. Below, we give a brief statement on each of the sources of potential systematic effects and refer to the pre-launch papers for a deeper analysis [32,34]:

\begin{itemize}
    \item \textbf{Instrumental and extra-galactic backgrounds.} From the stacking of almost 2 million seconds worth of background-rejected data from annular background regions around extra-galactic point sources (see Tab.~\ref{tab:bkgobservations}, we also excluded sources bright enough to pollute the background extraction region with their signal because of the wing of the point-spread-function), the background is unpolarized down to a minimum detectable polarization of 3.7\% at 99\% confidence level. This is well below the measured polarization from the Sgr~A region presented in this work.
    \item \textbf{Spurious polarization.} The spurious polarization we managed to calibrate in a 4.5' radius amounts for 0.05\%. The statistical error on the derived polarization measurement is 11\% (or about 4\% modulation). As a consequence, the residual systematic polarization has no effect on the results presented in this paper.
    \item \textbf{Stray light.} The IXPE requirement on the level of stray light is that its suppression should be larger than a factor of 200 (verified at 2.3 keV). We re-scaled the suppression coefficient of the three more intense sources outside the 4.5' radius circle used in this paper as found in the Chandra catalogue (see Tab.~\ref{Tab:Sources}). The resulting stray-light counting rate (0.5--7 keV) from these sources outside the field of view are 10 $\mu$Crab at most (Crab 0.5--7 keV = 2.8 $\times$ 10$^{-8}$ erg~s$^{-1}$~cm$^{-2}$). Assuming that these sources are 10\% polarized, the effect on the observed counting rate is at most 0.1\%, i.e. negligible.
    \item \textbf{Solar Effects.} Due to the thick shielding of the detectors, only in case of a very bright X Class solar flare a few photons have been detected (cross-matched by monitoring satellites). Data in coincidence of solar flares are simply removed. In addition, spikes due to passages over the SAA were also removed, as stated previously, and thus do not contribute to systematic effects.
    \item \textbf{Effects related to Earth atmosphere.} The Earth atmosphere, during most of observations, crosses the field of view of IXPE at each orbit. The data are excluded from the analysis. In any case, the atmosphere is dark in the range of IXPE and becomes bright above 10 keV. So we can state that the albedo of Earth is never directly observed by IXPE telescopes. Indirectly, namely through scattering, it contributes to the instrumental background. In fact the satellite orbits the Earth with an angular velocity of 360$^\circ$ in 90 minutes, which is 4$^\circ$ per minute. This means that the source is seen on the limb of the Earth only within 13', which is 3 seconds each orbit.
\end{itemize}

\subsection*{Chandra data reduction}

The Chandra data reduction follows a standard procedure based on the latest versions of the data reduction software (CIAO v.\ 4.14) and calibration (CALDB v.\ 4.9.8). Our particular approach and analysis steps are described in detail in [35]. Briefly, they include identification and removal of high background periods, correction of photon energies for the time and detector temperature dependence of the charge transfer inefficiency and gain, and creation of matching background datasets using blank sky observations with exposure times similar to the Galactic Center pointings.

The Chandra observing program included multiple pointings with individual exposures 10--30~kiloseconds. Due to the spacecraft orientation relative to the Sun, many observations were done with the ACIS focal plane temperature significantly higher than a nominal value of $-120$\,C, resulting in potential problems with the detector gain and response calibration. To minimize the impact of this issue, we have used three Chandra pointings (OBSIDs 24373, 24820, and 26353) with the lowest focal plane temperatures ($T<-116.5$\,C) and that were done quasi-simultaneously with the IXPE observations (February 26 through March 11), with a total exposure of 52~kiloseconds. 

For the analysis presented here, we use the combined flat-fielded and background-subtracted Chandra image in the 4--8~keV band, and spectra extracted in $4.5'$ circular region discussed above. Following the standard approach for analysing Chandra spectra of extended sources, we have generated the spectral response files that combine the position-dependent ACIS calibration with the weights proportional to the observed brightness in the 4--8 keV band. The extracted spectral data and response files are fit to a set of models as described below.

To assess the impact of the calibration uncertainties associated with the elevated focal plane temperature, we repeated all analyses using the OBSIDs with higher temperatures than the three pointings used in this work. These experiments have shown that after the standard temperature-dependent corrections implemented in CIAO, the parameters of interest for this work (e.g., the 4--8 keV flux of the reflected component) are almost unchanged. Therefore, we conclude that the impact of elevated focal plane temperatures on our analysis is negligible.

\subsection*{Archival observations with XMM-Newton}
In order to better constrain spectral de-composition of the X-ray emission from the region of interest, we have used all archival observations with XMM-Newton observatory accumulated between 2002 and 2012. These data have already been used for studies of the reflection component [36]. The XMM-Newton data were prepared, reduced and processed in exactly the same manner. As the final step, the spectrum was extracted from the same region as used by IXPE. Thanks to large effective area above 4 keV and long total exposure time supplemented with superior spectral resolution around 6 keV, the XMM-Newton data allow us to better separate thermal and reflected components. Since only the contribution of the reflected component is expected to vary over time, one can combine the archival data with the quasi-simultaneous Chandra and IXPE observations by allowing only parameters of the reflected component model to vary while linking parameters of all other components when fitting the spectra of the total emission observed by all three observatories.

\subsection*{Combined analysis of IXPE, Chandra, and XMM-Newton data}

As the first step of the joint IXPE, Chandra and XMM-Newton analysis, the background-subtracted spectra extracted from the circular region shown in Fig.~\ref{Fig:IXPE_Chandra_map} have been fitted with the same multi-component model.

This model includes a reflection component and three ``thermal" components (an illustrative example of the spectral model is shown in Fig.~\ref{fig:spectral_model_demo}). The latter components can be approximated as emission of optically thin thermal plasma, even though they might now be indeed associated with diffuse X-ray emitting gas. Indeed, in the 4--8 keV energy band, the most important is the thermal component with temperature $kT\sim 6\,{\rm keV}$, which is plausibly due to cumulative emission of large number of point sources, mostly accreting white dwarfs and stars with active coronae [37]. The \texttt{APEC} model (see \url{http://www.atomdb.org/}) was used for the optically thin thermal components. The reflection component describes the spectrum emerging from a spherical cloud of molecular gas, illuminated by a beam of X-rays, having a power-law spectrum. For this purpose, a publicly available \texttt{CREFL16} model [36] was used, and more specifically a version of it where reflected continuum and fluorescent lines are treated separately but with tied parameters. \texttt{CREFL16} is a model describing the spectrum of a uniform gas cloud illuminated by a parallel beam of X-rays. The model covers the energy range from 0.3 to 100 keV. It has 5 parameters: the Thomson optical depth of the cloud, the photon index of the incident spectrum, the abundance of heavy elements, the cosine of the angle between the line of sight and the illuminating angle, and the normalization. The softer thermal components are needed to properly describe the line-rich spectrum at energies below 3-4~keV (see Fig.~\ref{Fig:IXPE_Chandra_spectra}). Similarly to the hotter component discussed above, the exact nature of these components is still a matter of debate [38].

A combination of all components provides a reasonably good approximation of the IXPE, Chandra, and archival XMM-Newton spectra extracted from the reference region. XSPEC version 12.10.1f was used for fitting the spectra of Chandra, XMM-Newton, and IXPE. For IXPE, XSPEC was also used to fit the Q and U spectra too. We note in passing that IXPE and Chandra observations are quasi-simultaneous, while the XMM-Newton data were averaged over several observations spread over many years (from 2002 to 2012). For that reason, the normalization of the reflected component in the XMM-Newton model was decoupled from the IXPE and Chandra models. All other components/parameters were tied across all three instruments. A standard $\chi^2$ minimization has been used to fit all spectra (polarized and total). This approach is suitable for the IXPE, XMM-Newton, and Chandra data sets used here. The best-fitting spectral model (Fig.~\ref{Fig:IXPE_Chandra_spectra}) to the combined data set of three instruments  has a $\chi^2=1400$ for 922 degrees of freedom. the residuals shown in the second panel of Fig.\ref{Fig:IXPE_Chandra_spectra} is of the order of $\sim$20\%, providing a reasonable approximation of the data. The excess of the $\chi^2$ above expectations ($\chi^2=922\pm 43$ at the 68\% confidence) for the genuine model is not surprising given the simplicity of the model used here and, also, modest cross-calibration issues among all three telescopes.

The outcome of the spectral fitting model was in particular the normalization of the reflected component (expressed in units of surface brightness) and the parameters of the best-fitting \texttt{CREFL16} model, which do not affect significantly the results of the IXPE data analysis. 
Critically assessing the spectral decomposition shown in Fig.~\ref{Fig:IXPE_Chandra_spectra}, we concluded that a systematic uncertainty of order 30\% is associated with the derived normalization of the reflected component.  Since only the scattered continuum of the reflected component is polarized (ignoring second order effects caused by multiple scatterings, [39,40]), it was singled out and used to fit Q and U spectra measured by IXPE as shown in Fig.~\ref{Fig:IXPE_Q_U}. 
For that fit, the scattered continuum component was multiplied by the effective area appropriate for the extraction region (taking into account vignetting) and the modulation factor. For Q and U spectra, the dependence on the polarization angle $\phi$ was accounted for by introducing multiplicative factors $\cos 2\phi$ and $\sin 2\phi$ in front of the scattered continuum component, for Q and U, respectively. Namely, for the Thomson scattering,
\begin{eqnarray}
Q(E)=P\cos 2\phi \times S(E) \nonumber \\
U(E)=P\sin 2\phi \times S(E), \nonumber
\end{eqnarray}
where $S(E)$ is the spectrum of the scattered continuum as derived from the spectral analysis and $P$ is the degree of polarization.
This model is valid in the single scattering approximation, and for a compact scattering cloud. From these expressions, it is clear that $P$ is degenerate with the normalization of $S(E)$, but it does not affect the detection significance of non-zero polarization. This allows one to use a multiplicative notation when quoting the polarization degree as $P=(31\pm11)\% \times (1\pm 0.3)$, where the second term arises from the uncertainties in the reflected continuum normalization. 

The best-fitting model to the Q and U spectra (Fig.~\ref{Fig:IXPE_Q_U}) has a $\chi^2=254.4$ for 248 unbinned channels and two free parameters - the normalization of the polarized component and the polarization angle. In the figure, the channels were grouped (12 channels per bin) for display purposes.

\subsection*{Sgr~A$^*$ flare parameters}
\
More than a century ago, Kapteyn [41] realized that for a flare of a compact source and an arbitrary observer, its scattered light with a given time delay since the moment of the flare is associated with the matter located on the surface of an ellipsoid of rotation. The source and the observer are at the foci of the ellipsoid. Furthermore, in close vicinity of the source, the ellipsoid can be approximated by a paraboloid. Assuming that Sgr~A$^*$ is the primary source, a detection of the scattered light from a cloud relates the age of the flare and the position of the cloud as
\begin{eqnarray}
t=\frac{\sqrt{x^2+z^2}}{c}+\frac{z}{c},
\end{eqnarray}
where $c$ is the speed of light and $x$ and $z$ are the distances of the cloud from Sgr~A$^*$ in the sky plane and along the line of sight, respectively. 

The degree of polarization $P$ depends on the cosine of the scattering angle $\mu$ 
(in the limit of Thomson scattering) as
\begin{eqnarray}
P=\frac{1-\mu^2}{1+\mu^2},
\end{eqnarray}
where
\begin{eqnarray}
\mu=-\frac{z}{\sqrt{x^2+z^2}}.
\end{eqnarray}
Thus, knowing $P$, one can reconstruct the 3D position of illuminated clouds. For instance, for $x=25\,{\rm pc}$ and $\theta=137^\circ$, $z\approx 26 \,{\rm pc}$, i.e. the illuminated clouds are further away from us than Sgr~A$^*$ by this distance. 
Combining the above three equations, one can explicitly relate the delay time $t$ and the measured degree of polarization $P$
\begin{eqnarray}
t=\frac{x}{c}\times \left [ \left ( \frac{1+P}{2P} \right )^{1/2} \pm \left ( \frac{1-P}{2P} \right )^{1/2}\right].
\end{eqnarray}
This expression was used to estimate the age of the flare.

The determination of the flare luminosity is notoriously difficult, since it depends not only on the distance of the cloud from  Sgr~A$^*$, but also on the density of the cloud gas and the duration of the flare. One can partly circumvent this problem by using the maximal possible albedo of a molecular cloud [36] to convert the observed X-ray surface brightness into a lower limit on the luminosity
\begin{eqnarray}
L_{4-8,{\min}}=\frac{I_{4-8}}{\eta_{\max}} 4\pi D_{\rm sc}^2\approx 8\times10^{37}\,{\rm erg\,s^{-1}}, 
\end{eqnarray}
where $I_{4-8}=4.6\times 10^{-13}\,{\rm erg\,s^{-1}\,cm^{-2}\,arcmin^{-2}}=5.4\times10^{-6}\,{\rm erg\,s^{-1}\,cm^{-2}\,sr^{-1}}$ is the observed flux from the studied region [27], $\eta_{\max}\approx 10^{-2}$ is the maximal albedo, and $D_{\rm sc}=\sqrt{x^2+z^2}\approx 36\,{\rm pc}$ is the distance between  Sgr~A$^*$ and the clouds. Conversion of $L_{4-8,{\min}}$ to a broader 1--100 keV band assuming a power-law spectrum with the photon index of 2.0 (the value typically measured for the Sgr~A GMCs [42]), yields  $L_{1-100,{\min}}\approx 6\times10^{38}\,{\rm erg\,s^{-1}}$. If the optical depth or, equivalently, the hydrogen column density of illuminated clouds is known, one can use the real albedo in the above equation, driving the minimum luminosity up. Similarly, if only a fraction of the area of the studied region is covered by illuminated clouds, the true luminosity will be higher. For instance, focusing on individual bright clouds, drives this estimate up by an order of magnitude to the level of a few $10^{39}\,{\rm erg\,s^{-1}}$ [36], and this is still a lower limit. 

However, there is no easy way of placing an upper limit on the source luminosity (for a short flare) without additional explicit assumptions on the gas density and the duration of the flare. For instance, an hour-long flare with a luminosity of $\sim 10^{44} {\rm erg~s^{-1}}$ could be consistent with the data.  

Finally, we note that the above derivation assumes that all visible reflection emission is due to a single flare. If Sgr~A$^*$ produced multiple flares over the past several hundred years [43,44] then the fluxes and polarization will be affected. Longer IXPE observations should be able to test this scenario by providing a polarization measurement of individual GMCs. 

\clearpage

\subsection*{References}

[27] Ildar Khabibullin, Eugene Churazov, and Rashid Sunyaev. SRG/eROSITA view of X-ray reflection in the Central Molecular Zone: a snapshot in September-October 2019. Monthly Notices of the Royal Astronomical Society, 509(4):6068–6076, February 2022.  \newline \newline
[28] Martin C. Weisskopf and et al. The Imaging X-Ray Polarimetry Explorer (IXPE): Pre-Launch. Journal of Astronomical Telescopes, Instruments, and Systems, 8(2):026002, April 2022. \newline \newline
[29] Riccardo Ferrazzoli and et al. In-flight calibration system of imaging x-ray polarimetry explorer. Journal of Astronomical Telescopes, Instruments, and Systems, 6:048002, October 2020. \newline \newline
[30] John Rankin and et al. An Algorithm to Calibrate and Correct the Response to Unpolarized Radiation of the X-Ray Polarimeter Onboard IXPE. Astronomical Journal, 163(2):39, February 2022. \newline \newline
[31] F. Kislat, B. Clark, M. Beilicke, and H. Krawczynski. Analyzing the data from X-ray polarimeters with Stokes parameters. Astroparticle Physics, 68:45–51, August 2015. \newline \newline
[32] A. Di Marco and et al. The ixpe x-ray observatory background. In preparation, 2021. \newline \newline
[33] Luca Baldini and et al. ixpeobssim: A simulation and analysis framework for the imaging X-ray polarimetry explorer. SoftwareX, 19:101194, July 2022. \newline \newline
[34] Fabio Muleri and et al. Performance of the Gas Pixel Detector: an x-ray imaging polarimeter for upcoming missions of astrophysics. In Jan-Willem A. den Herder, Tadayuki Takahashi, and Marshall Bautz, editors, Space Telescopes and Instrumentation 2016: Ultraviolet to Gamma Ray, volume 9905 of Society of Photo-Optical Instrumentation Engineers (SPIE) Conference Series, page 99054G, July 2016. \newline \newline
[35] A. Vikhlinin and et al. Chandra Cluster Cosmology Project. II. Samples and X-Ray Data Reduction. Astrophysical Journal, 692(2):1033–1059, February 2009. \newline \newline
[36] E. Churazov, I. Khabibullin, G. Ponti, and R. Sunyaev. Polarization and long-term variability of Sgr~A$^*$ X-ray echo. Monthly Notices of the Royal Astronomical Society, 468(1):165–179, June 2017. \newline \newline
[37] M. Revnivtsev and et al. Discrete sources as the origin of the Galactic X-ray ridge emission. Nature, 458(7242):1142–1144, April 2009. \newline \newline
[38] Takayuki Yuasa, Kazuo Makishima, and Kazuhiro Nakazawa. Broadband Spectral Analysis of the Galactic Ridge X-Ray Emission. Astrophysical Journal, 753(2):129, July 2012. \newline \newline
[39] E. Churazov, R. Sunyaev, and S. Sazonov. Polarization of X-ray emission from the Sgr B2 cloud. Monthly Notices of the Royal Astronomical Society, 330(4):817–820, March 2002. \newline \newline
[40] F. Marin, V. Karas, D. Kunneriath, and F. Muleri. Prospects of 3D mapping of the Galactic Centre clouds with X-ray polarimetry. Monthly Notices of the Royal Astronomical Society, 441(4):3170–3176, July 2014. \newline \newline
[41] J. C. Kapteyn. On the motion of Nebulae in vicinity of Nova Persei. Popular Astronomy, 10:124–127, March 1902. \newline \newline
[42] G. Ponti, R. Terrier, A. Goldwurm, G. Belanger, and G. Trap. Discovery of a Superluminal Fe K Echo at the Galactic Center: The Glorious Past of Sgr~A$^*$ Preserved by Molecular Clouds. Astrophysical Journal, 714(1):732–747, May 2010. \newline \newline
[43] M. Clavel and et al.. Echoes of multiple outbursts of Sagittarius A$^*$ revealed by Chandra. Astronomy and Astrophysics, 558:A32, October 2013. \newline \newline
[44] D. Chuard and et al. Glimpses of the past activity of Sgr~A$^*$ inferred from X-ray echoes in Sgr C. Astronomy and Astrophysics, 610:A34, February 2018. \newline \newline
[45] L. Sidoli and S. Mereghetti. The X-ray diffuse emission from the Galactic Center. Astronomy and Astrophysics, 349:L49–L52, September 1999. \newline \newline
[46] L. Di Gesu and et al. Prospects for IXPE and eXTP polarimetric archaeology of the reflection nebulae in the Galactic center. Astronomy and Astrophysics, 643:A52, November 2020.

\end{document}